\documentclass[11pt]{article}
\usepackage{amsmath,amssymb,graphics,graphicx,amsthm}
\usepackage{comment,url,enumitem}
\usepackage{xspace}

\newtheorem{theorem}{Theorem}[section]

\begin{document}

\newcommand{\EM}{${\cal E}_M$\xspace}
\newcommand{\Em}{{\cal E}_M}
\newcommand{\MM}{$\cal M$\xspace}
\newcommand{\cc}{{\em count}\xspace}
\newcommand{\As}{${\cal A}_s$\xspace}
\newcommand{\IMv}{{\sc In-Mis}$(v)$\xspace}

\date{}
\title{Simple dynamic algorithms for Maximal Independent Set\\ and other problems}

\author{Manoj Gupta\\ IIT Gandhinagar, \\Gandhinagar, India\\
  \texttt{gmanoj@iitgn.ac.in} \and 
  Shahbaz Khan\thanks{
This research work was supported by the 
European Research Council under the European Union's Seventh Framework 
Programme (FP/2007-2013)~/~ERC Grant Agreement no. 340506.}\\
  Faculty of Computer Science, \\University of Vienna, Austria\\
  \texttt{shahbaz.khan@univie.ac.at}}
  
%



\maketitle
\begin{abstract}

Most graphs in real life keep changing with time. These changes can be in the form of insertion or deletion of edges or vertices. Such rapidly changing graphs motivate us to study dynamic graph algorithms. However, three important graph problems that are perhaps not sufficiently addressed in the literature include independent sets, maximum matching (exact) and maximum flows.

Maximal Independent Set (MIS) is one of the most prominently studied problems in the distributed setting. Recently, the first dynamic MIS algorithm for distributed networks was given by Censor-Hillel et al.~[PODC16], requiring expected $O(1)$ amortized rounds with $O(\Delta)$ messages per update, where $\Delta$ is the maximum degree of a vertex in the graph. They suggested an open problem to maintain MIS in fully dynamic centralized setting more efficiently. Assadi et al.~[STOC18] presented a deterministic centralized fully dynamic MIS algorithm requiring $O(\min\{\Delta,m^{3/4}\})$ amortized time per update. 
This result is quite complex involving an exhaustive case analysis. We report a surprisingly simple deterministic {\em centralized} algorithm which improves the amortized update time to $O(\min\{\Delta,m^{2/3}\})$. 

Additionally, we present some other minor results related to dynamic MIS, Maximum Flow, and Maximum Matching. 
A common trait of all our results is that despite improving state of the art upper bounds or matching state of the art lower bounds, they are surprisingly simple and are analysed using simple amortization arguments. Further, they use no complicated data structures or black box algorithms for their implementation.

\end{abstract}

\pagenumbering{gobble}
\newpage
\pagenumbering{arabic} 
\setcounter{page}{1}
\section{Introduction}
In the last two decades, there has been a flurry of results in the area of dynamic graph algorithms. The motivation behind studying such problems is that many graphs encountered in the real world keep changing with time. These changes can be in the form of addition and/or deletion of edges and/or vertices. The aim of a dynamic graph algorithm is to report the solution of the concerned graph problem after every such update in a graph. One can trivially use the best static algorithm to compute the solution from scratch after each such update. Hence, the aim is to update the solution more efficiently as compared to the best static algorithm. Various graph problems studied in the dynamic setting include connectivity~\cite{HenzingerK99,HolmLT01,KapronKM13}, minimum spanning tree~\cite{EppsteinGIN97,Frederickson85,NanongkaiSW17}, reachability~\cite{DemetrescuI08,Sankowski04,RodittyZ16}, shortest path~\cite{Thorup05,BaswanaHS07,DemetrescuI04}, matching \cite{Bhattacharya2017b,BaswanaGS15,GuptaP13}, etc. However, three important graph problems that are perhaps not sufficiently addressed in the literature include independent sets, maximum matching (exact) and maximum flows.

For a given graph $G=(V,E)$ with $n$ vertices and $m$ edges where maximum degree of a vertex is $\Delta$, a set of vertices \MM$\subseteq V$ is called an {\em independent set} if no two vertices in \MM share an edge in $E$, i.e. $\forall x,y\in$\MM, $(x,y)\notin E$. Computing the maximum cardinality {\em independent set} is known to be NP-Hard~\cite{GareyJ90}. However, a simple greedy algorithm is known to compute the Maximal Independent Set (MIS) in $O(m)$ time, where MIS is any independent set \MM such that no proper superset \MM$'$ of \MM, i.e. \MM$\subset$\MM$'$, is an independent set of the graph. Note that MIS is not a good approximation of maximum cardinality independent set\footnote{Consider a star graph where a vertex $s$ has an edge to all other vertices in $V$. The maximum independent set is \MM$^*=V\setminus\{s\}$, whereas \MM$'=\{s\}$ is a valid MIS. Here approximation factor $|$\MM$^*|/|$\MM$'|$ is $n-1$.} unlike other known problems such as matching (which is a 2-approximation of maximum matching).

\subsection*{Fully Dynamic Maximal Independent Set}
In the dynamic setting, one can trivially maintain MIS in $O(m)$ update time, by computing MIS from scratch after every update. Moreover, the {\em adjustment} complexity\footnote{The {\em adjustment} complexity refers to the number of vertices that enter or leave \MM after an update.} of this algorithm can be $O(n)$. Until recently no non-trivial algorithm was known to maintain MIS in $o(m)$ time.

The problem of MIS has been extensively studied in the distributed setting~\cite{BarenboimEPS12,Ghaffari16,JeavonsS016}. The first dynamic MIS algorithm was given by Censor-Hillel et al.~\cite{CensorHK16} showing the maintenance of dynamic MIS in expected $O(1)$ amortized rounds with $O(\Delta)$ messages per update, where $\Delta$ is the maximum degree of a vertex in $G$. This also translates to a {\em centralized} algorithm with update time of amortized $\Omega(\Delta)$~\cite{AssadiOSS18}.  Recently, Assadi et al.~\cite{AssadiOSS18} presented a deterministic {\em centralized} fully dynamic MIS algorithm requiring $O(\min\{\Delta,m^{3/4}\})$ amortized time and $O(1)$ amortized adjustments per edge update. Further, the update time for centralized algorithm was recently improved~\cite{OnakSSW18} for low arboricity graphs. In this paper, we present surprisingly simple {\em deterministic} algorithms to improve the results for general arboricity graphs.

\begin{theorem}[Fully dynamic MIS (centralized)]
Given any graph $G=(V,E)$ having $n$ vertices and $m$ edges, MIS can be maintained in $O(\min\{\Delta,m^{2/3}\})$ amortized time and $O(1)$ amortized adjustments per insertion or deletion of an edge or a vertex, where $\Delta$ is the maximum degree of a vertex in $G$.
\label{thm:fdMIS}
\end{theorem}

\noindent
\textbf{Remark: } A limitation of our result over Assadi et al.~\cite{AssadiOSS18} is that it cannot be trivially extended to the distributed setting with efficient round and adjustment complexity. Trivial adaptation of our algorithm in the distributed ${\cal CONGEST}$ model requires $O(m^{1/3})$ amortized rounds and adjustments per update, whereas that of \cite{AssadiOSS18}  requires $O(1)$ rounds and adjustments per update. The message complexity in both cases matches the update time of the centralized setting. 


Additionally, we present some other minor results related to dynamic MIS, Maximum Flow, and Maximum Matching as follows. 
\begin{enumerate}

\item \textbf{Hardness of dynamic MIS and Incremental MIS}\\
We first discuss the hardness of different dynamic updates for maintaining MIS and report that the hardest update for maintaining MIS is that of edge insertions. For the remaining dynamic settings (fully dynamic vertex updates and decremental edge updates), a simple algorithm~\cite{AssadiOSS18} maintains MIS optimally, i.e. total update time is of the order of input size. Hence, in addition to fully dynamic edge updates, the only interesting dynamic setting is handling edge insertions, where we again present a very simple algorithm to prove the following.

\begin{theorem}[Incremental MIS]
Given any graph $G=(V,E)$ having $n$ vertices and $m$ edges, MIS can be maintained in $O(\min\{\Delta,\sqrt{m}\})$ amortized time and $O(1)$ amortized adjustments per edge insertion, where $\Delta$ is the maximum degree of a vertex in $G$.
\label{thm:incMIS}
\end{theorem}

\noindent
\textbf{Remark:} Both these algorithms (for incremental MIS and remaining simple updates) trivially extend to the distributed {\cal CONGEST} model with $O(1)$ amortized rounds per update.

\item \textbf{Worst case bounds for dynamic MIS}\\
All the previous results for dynamic MIS primarily focus on amortized guarantees. This was explained by Assadi et al.~\cite{AssadiOSS18} by proving that the {\em adjustment} complexity (and hence update time) of any dynamic MIS algorithm must be $\Omega(n)$ in the worst case. Hence, in order to achieve better worst case bounds, we are required to consider a {\em relaxed} model. Thus, we relax the requirement to {\em explicitly} maintain the MIS after each update. Rather, we allow queries of the form \IMv, which reports whether a vertex $v\in V$ is present in some MIS of the updated graph, ensuring the following properties. Firstly, the responses to all the queries after an update are consistent to some MIS of the updated graph. Secondly, the adjustment complexity of the algorithm is amortized $O(1)$ per update (considering only updates), or worst case $O(1)$ per update and query. Such a model have been previously studied for several problems including MIS~\cite{RubinfeldTVX11,AlonRVX12,MansourRVX12}.  In this {\em relaxed model} we show that 
\begin{theorem}[Fully dynamic MIS (worst case)]
Given any graph $G=(V,E)$ having $n$ vertices and $m$ edges, MIS can be implicitly maintained under fully dynamic edge updates requiring $O(1)$ adjustments per update and query, which allows queries of the form \IMv, where both update and query require worst case $O(\min\{\Delta,\sqrt{m}\})$ time.
\end{theorem}

\noindent
\textbf{Remark:} It trivially extends to {\cal CONGEST} model with $O(1)$ rounds per update and query.

\item \textbf{Dynamic Maximum Flow and Maximum Matching}\\
The maximum flow and maximum matching problems are some of the most studied combinatorial optimization problem having a lot of practical applications. Recently, Dahlgaard \cite{Dahlgaard16} proved conditional lower bounds for partially dynamic problems including maximum flow and maximum matching. Assuming the correctness of OMv conjecture, he proved that maintaining incremental Maximum Flow or Maximum Matching for unweighted graphs requires $\Omega(n)$ update time (even amortized).  We report trivial extensions of two classical  algorithms based on {\em augmenting paths}, namely incremental reachability algorithm~\cite{Italiano86} and blossoms algorithm~\cite{Edmonds87,Gabow74}, which match these lower bounds. 

For the sake of completeness, we also report the folklore algorithms to update the maximum flow of an unweighted (unit-capacity) graph and maximum cardinality matching, in $O(m)$ time using simple reachability queries. To the best of our knowledge, it is widely known but so far it has not been a part of any literature. 
\end{enumerate}

\section{Overview}
Let the given graph be $G=(V,E)$ with $n$ vertices and $m$ edges, where the maximum degree of a vertex is $\Delta$. The degree of a vertex $v\in V$ shall be denoted by $deg(v)$. We shall represent the currently maintained MIS by $\cal M$. We shall now give a brief overview of our results and the difference of our approach from the current state of the art.

A trivial static algorithm can compute an MIS of a graph in $O(m)$ time. It visits each vertex $v$ and checks whether any of its neighbours is in \MM. If no such neighbour exists, it adds $v$ to \MM, clearly taking $O(1)$ time for each edge while visiting its endpoints.

Assadi et al.~\cite{AssadiOSS18} described a {\em simple} dynamic algorithm (henceforth referred as \As) requiring $O(\Delta)$ amortized time per update, where essentially each vertex maintains a \cc of the number of its neighbours in \MM. They also described an improved algorithm requiring $O(\min\{\Delta, m^{3/4}\})$ time, which is based on a complicated case analysis. The algorithm essentially divides the vertices into four sets based on their degrees, where the \cc of all the vertices except the low degree vertices $V_{low}$ is maintained exactly. For vertices in $V_{low}$, instead some {\em partial estimate} of \cc is maintained ignoring the high degree vertices in \MM. 
The key difference of this improved algorithm from \As is the following: Instead of adding a vertex to \MM only when none of its neighbours is in \MM (i.e. when the true value of \cc is zero), in some cases the algorithm adds a low degree vertex to \MM even when merely the {\em partial estimate} of \cc is zero. As a result, the algorithm may need to remove some vertices from \MM to ensure correctness of the algorithm. 

Now, the simple \As algorithm can be used to maintain MIS under all possible graph updates. We firstly provide a {\em mildly} tighter analysis of \As to demonstrate what kind of dynamic settings are harder for the MIS problem. We show that \As solves the problem optimally for all kinds of updates except {\em edge insertions}. Hence, the two dynamic settings in which the MIS problem is interesting are the fully dynamic and the incremental settings under {\em edge updates}. We improve the state-of-the-art for dynamic MIS in both these settings using very simple algorithms, which are essentially based on the simple \As algorithm as follows.

Our {\em fully dynamic algorithm} processes the low degree vertices $V_{low}$ (say having degree $\leq \Delta_c$) totally independent of the high degree vertices. Note that this simplifies the approach of \cite{AssadiOSS18} since the key difference of their algorithm from \As is the partial independence of processing $V_{low}$ with respect to higher degree vertices. Hence, we maintain the MIS \MM of the subgraph induced by the vertices in $V_{low}$ irrespective of its high degree neighbours in \MM. Using \As the MIS of this subgraph can be maintained in $O(\Delta_c)$ amortized update time. After each update, the MIS of high degree vertices not adjacent to any low vertex in \MM (i.e. \MM$\cap~ V_{low}$) can be recomputed from scratch using the trivial static algorithm. This gives a trade off since the size of the subgraph induced by the high degree vertices (and hence the time taken by the static algorithm) decreases as $\Delta_c$ is increased. Hence, choosing an appropriate $\Delta_c$ results in amortized $O(\min\{\Delta,m^{2/3}\})$ update time. Note that recomputing the MIS for high degree vertices from scratch may lead to a larger {\em adjustment} and {\em round} complexity in the distributed setting.




In the {\em incremental setting}, we show that \As indeed takes $\Theta(\Delta)$ amortized time per update (see Appendix~\ref{sec:tightInc}). Moreover, we improve \As using a simple modification which prioritizes the removal of low degree vertices from \MM instead of an arbitrary choice by \As. This simple modification improves the amortized update time to $O(\min\{\Delta, \sqrt{m}\})$, which is also shown to be tight (see Appendix~\ref{sec:tightInc}).

For {\em worst case complexity} of a fully dynamic MIS algorithm Assadi et al.~\cite{AssadiOSS18} showed strong lower bounds, where a single update may lead to $\Theta(n)$ vertices to enter or leave \MM. Hence, for achieving better {\em worst case} bounds our relaxed model essentially maintains merely an {\em independent set} explicitly rather than MIS. Thus, the MIS is completely built over several queries allowing better worst case complexity. We present a simple algorithm for maintaining fully dynamic MIS in this model requiring $O(\min\{\Delta,\sqrt{m}\})$ worst case time for update and query. 

\section{Dynamic MIS}
Assadi et al.~\cite{AssadiOSS18} demonstrated a simple algorithm \As for maintaining fully dynamic MIS using $O(\Delta)$ amortized time per update. We first briefly describe the algorithm and its analysis, which shall be followed by a tighter analysis that can be used to argue the kind of dynamic updates for which it is harder to maintain MIS. 
 
\subsection{Simple dynamic algorithm \As~\cite{AssadiOSS18}}
The following algorithm is the most natural approach to study the dynamic MIS problem. It essentially maintains the {\em count} of the number of neighbours of a vertex in the MIS. It is easy to see that the {\em count} for each vertex can be initialized in $O(m)$ time using the simple greedy algorithm for computing the MIS of the initial graph. 

Now, under dynamic updates the \cc of each vertex needs to be maintained explicitly. Hence, whenever a vertex $v$  enters or leaves \MM, the \cc of its neighbours is updated in $O(deg(v))$ time. On insertion of a vertex $v$, the \cc of $v$ is computed in $O(deg(v))$ time. In case this \cc is {\em zero}, $v$ is added to \MM. In case of deletion of a vertex $v$, an update is only required when $v\in$\MM, where $v$ is simply removed from \MM. In case \cc of any neighbour of $v$ reduces to {\em zero}, it is added to \MM. In case of deletion of an edge $(u,v)$, update is required only when one of them (say $u$) is in \MM. If \cc of $v$ reduces to {\em zero}, it is added to \MM. Finally, on insertion of an edge $(u,v)$, update is required only in case both are in \MM, where either one of them (say $v$) is removed from \MM. Again, if \cc of any neighbour of $v$ reduces to {\em zero}, it is added to \MM.

Notice that in case of each update at most {\em one} vertex may be removed from \MM and several vertices may be added to \MM, where both addition or deletion of a vertex $v$ takes $O(deg(v))$ time. Hence, whenever a vertex is removed from \MM, $O(1)$ adjustments and $O(\Delta)$ work is charged for both this removal as well as the next insertion (if any) to \MM. The initial charge required is the sum of degrees of all the vertices in \MM, which is $O(m)$. As a result, fully dynamic MIS can be maintained in $O(1)$ amortized adjustments and $O(\Delta)$ amortized time per update.

\begin{theorem}[Fully dynamic MIS~\cite{AssadiOSS18}]
Given any graph $G=(V,E)$ having $n$ vertices and $m$ edges, MIS can be maintained in $O(1)$ amortized adjustments and $O(\Delta)$ amortized time per insertion or deletion of edge or vertex, where $\Delta$ is the maximum degree of a vertex in $G$.
\end{theorem}

\subsubsection{Tighter analysis}
We shall now show a mildly tighter analysis of the algorithm which is again very simple and apparent from the algorithm itself. Instead of generalizing the update time to $O(\Delta)$, we show that the update time of the algorithm is $O(deg(v))$, where $v$ is the vertex that is removed from \MM or inserted in the graph (for vertex insertions), otherwise the charged update time is $O(1)$. 

This analysis is again fairly straight forward using the similar arguments as in the previous section, which stated that a vertex can be charged twice $O(\Delta)$ when it is removed from \MM to pay for the future cost of its insertion to \MM. The only difference in our analysis is as follows: Since we are not charging the update with the maximum degree $\Delta$, rather the exact degree $deg(v)$ of a vertex $v$, this $deg(v)$ may change between the time it was charged and when it is used. Particularly, consider a vertex $v$ which was charged $deg(v)$ when removed from \MM or inserted in the graph. Now, on being inserted back in \MM, its new $deg(v)$ can be much higher than its old $deg(v)$. Hence, we need to explicitly associate this increase of degree to the update which led to this increase. To this end, we  analyze the algorithm using the following potential function: $\Phi = \sum_{v\notin \text{\MM}} deg(v)$.  Thus, the amortized cost (time) of an update is the sum of the actual work done and the change in potential.

The insertion of a vertex $v$ requires amortized $O(deg(v))$ time  for the actual $deg(v)$ work to add the edges, and the increase in potential $\Phi$. The potential $\Phi$ can be increased by $O(deg(v))$ because of $v$ (if $v\notin$\MM), and its neighbours not in \MM (change in their degrees). For the remaining updates, note that the work done for the addition of a vertex $v$ in \MM is always balanced by the corresponding decrease in potential by $deg(v)$. {\em We thus only focus on the work done for removing a vertex from \MM and the change in potential due to  change in the degree of vertices}. The deletion of a vertex $v$ requires amortized $O(1)$ time, as the  time required to remove  edges of $v$ is $deg(v),$ and the potential $\Phi$ decreases by at least $deg(v)$ (if $v\notin~$\MM  then $\Phi$ reduces due to $v$, else it reduces due to neighbours of $v$). The insertion of an edge $(u,v)$ requires  amortized $O(1)$ time if both $u$ and $v$ are not simultaneously in \MM, for the $O(1)$ work done to add the edge in the graph and $O(1)$ increase in $\Phi$ because of degrees of its endpoints. In case both $u,v\in$\MM, the algorithm removes one vertex (say $v$) from \MM, which requires amortized $O(deg(v))$ time for actual $deg(v)$ time taken to update all the neighbours of $v$ and increase in potential $\Phi$ by $O(deg(v))$, because of $v$ and its neighbours. The deletion of an edge $(u,v)$ again takes amortized $O(1)$ time for  $O(1)$ work done to remove the edge and  decrease in  $\Phi$. Finally, since at most one vertex (if any) is removed from \MM during an update, the adjustment complexity is amortized $O(1)$ as the removal also accounts for the future insertion of the vertex in \MM. Thus, we have the following theorem.

\begin{theorem}
Given any graph $G=(V,E)$ having $n$ vertices and $m$ edges, MIS can be maintained using $O(1)$ amortized adjustments per update, where the amortized update time is $O(deg(v))$ if $v$ is the only vertex (if any) that is removed from \MM or the vertex inserted into $G$ (vertex insertion), otherwise the amortized update time is $O(1)$. 
\label{thm:simpleAlg}
\end{theorem}

\noindent
\textbf{Remark:} This does not imply tighter amortized bound to the fully dynamic algorithm, but is merely described to aid in the analysis of future algorithms. Further, since adjustment complexity is amortized $O(1)$, the algorithm can be trivially adapted to the distributed $\cal CONGEST$ model~\cite{AssadiOSS18} requiring amortized $O(1)$ rounds per update.


\subsubsection{Hardness of Dynamic MIS}
Using Theorem~\ref{thm:simpleAlg} we can understand the hardest form of update in case of dynamic MIS. Additionally, we use the fact that a vertex $v$ is removed from $\cal M$ only in case of edge insertion or when $v$ is deleted. Consider the case of fully dynamic MIS under only vertex updates. Here each vertex $v$ can enter into \MM only once, either on being inserted or when all its neighbours in \MM are deleted. Thus, it requires total $O(deg(v))$ time throughout its lifetime, requiring overall $O(m)$ time. Similarly, consider the case of decremental MIS under edge updates. Here again, each vertex $v$ can enter \MM exactly once, and never leave \MM. Hence, using Theorem~\ref{thm:simpleAlg}, it incurs amortized cost of $O(1)$, requiring total $O(m)$ time. Thus, the algorithm \As works optimally under fully dynamic vertex updates or decremental MIS under edge updates. 

As a result, the only update for which the simple algorithm does not solve the problem optimally is that of edge insertions, leaving the two problems of incremental MIS and fully dynamic MIS under edge updates. Under the current analysis both these algorithm takes total $\Omega(m)$ time which may not be optimal. Thus, we shall now focus on solving the two problems better than the improved fully dynamic MIS algorithm by Assadi et al.~\cite{AssadiOSS18} requiring amortized $O(\min\{\Delta,m^{3/4}\})$ time.

\section{Improved algorithm for Fully Dynamic MIS}
\label{sec:fdMIS}
In addition to the simple $O(\Delta)$ amortized time algorithm \As, Assadi et al.~\cite{AssadiOSS18} presented a substantially complex fully dynamic algorithm requiring amortized $O(\min\{\Delta,m^{3/4}\})$ time per update. We show a very simple extension to \As that achieves amortized $O(\min\{\Delta,m^{2/3}\})$ time per update. The core idea used by the improved algorithm of Assadi et al.~\cite{AssadiOSS18} gives partial preference to low degree vertices when being inserted to \MM. More precisely, in {\em some cases} they allow a low degree vertex $v$ to be inserted to \MM despite having high degree neighbours in \MM. This is followed by removal of these high degree neighbours from \MM to maintain MIS property. We essentially use the same idea by using a stronger preference order. The low degree vertices are {\em always} inserted to \MM despite having high degree neighbours in \MM.

The main idea is as follows. We divide the vertices of the graph into $heavy$ vertices $V_H$ having degree $\ge \Delta_c$, and $light$ vertices $V_L$ having degree $<\Delta_c$, for some fixed constant $\Delta_c$. Let the subgraph induced by the vertices in $V_H$ and $V_L$ be $G_H=(V_H,E_H)$ and $G_L=(V_L,E_L)$ respectively. We use simple algorithm \As to maintain the MIS of $G_L$ in amortized $O(\Delta_c)$ time. After every such update, we can afford to rebuild the MIS for $G_H$ from scratch, using the trivial static algorithm. Since the total number of heavy vertices can be $O(m/\Delta_c)$, the time taken to build the MIS for the heavy vertices is   $O(|E_H|)=O((m/\Delta_c)^2)$ time. Choosing $\Delta_c=m^{2/3}$ we get an amortized $O(m^{2/3})$ update time algorithm. However, if $\Delta\leq m^{2/3}$ the entire graph is present in $G_L$ and we have an empty $G_H$. Hence, our algorithm merely performs \As on the whole graph, resulting in amortized $O(\min\{\Delta,m^{2/3}\})$ update time of our algorithm. Note that the main reason for faster update of this algorithm (compared to \As) is as follows: When a heavy vertex enters or leaves \MM, it does not inform its  light neighbours.

\subsubsection*{Implementation Details}
We shall now describe a few low level details regarding the implementation of the algorithm. Each vertex of both $V_H$ and $V_L$ will store a {\em count~} of their light neighbours in \MM, i.e. neighbours in $V_L~ \cap~ $\MM. This {\em count~} can be maintained easily by each vertex as whenever a light vertex enters of leaves \MM,  it informs all its $\Delta_c$ neighbours (both heavy and light).

As we have mentioned before, we use algorithm \As to maintain a maximal independent set for all the light vertices.  Thus, at each step we may have to rebuild the MIS for the heavy vertices from scratch. While rebuilding MIS for $G_H$ we  have to only consider those heavy vertices whose $count=0$, i.e., which do not have any light neighbour in \MM. We can rebuild the MIS of $G_H$ from scratch in $O(|E_H|)$ time using the trivial static algorithm. We have already argued that $|E_H| = O((m/\Delta_c)^2)$. Thus, the update time of our algorithm for light vertices is amortized $O(\Delta_c)$ (by algorithm \As) and the update time of our algorithm for heavy vertices is $O( (m/\Delta_{c})^2)$. Choosing $\Delta_c=m^{2/3}$, the update time of our algorithm becomes $O(m^{2/3})$.

Finally, note that the value of $\Delta_c$ have to be changed during the fully dynamic procedure, because of the change is the value of $m$. This can be achieved using the standard technique of {\em periodic rebuilding}, which can be described using {\em phases} as follows.  We choose $\Delta_c=m_c^{2/3}$, where $m_c$ denotes the number of edges in the graph at the start of  a phase. Hence at the start of the algorithm, $m_c=m$. Whenever $m$ decreases to $m_c/2$ or $m$ increases to $2m_c$, we end our current phase and start a new phase. Also, we re-initialize \cc and rebuild \MM  but using the new value of $m_c$, and hence new $V_L$. Note that the cost $O(m)$ for this rebuilding at the start of each phase is accounted to the $\ge m/2$ edge updates during the phase. Thus, we have the following theorem.

\newtheorem*{fdMIS}{Theorem~\ref{thm:fdMIS}}

\begin{fdMIS}
[Fully dynamic MIS]
Given any graph $G=(V,E)$ having $n$ vertices and $m$ edges, MIS can be maintained in $O(\min\{\Delta,m^{2/3}\})$ amortized time per insertion or deletion, where $\Delta$ is the maximum degree of a vertex in $G$.
\end{fdMIS}

\section{Improved Algorithm for Incremental MIS}
\label{sec:incMISsparse}
We shall now consider the incremental MIS problem and show that if we restrict the updates to edge insertion only, we can achieve a faster amortized update time of $O(\min\{\Delta, \sqrt{m}\})$. In this algorithm, we again give preference for a vertex being in MIS, based on its degrees. However, instead of dividing the vertices into sets of $V_L$ and $V_H$ explicitly, we simply consider the exact degree of the vertex.

Recall that in case of insertion of an edge, the simple algorithm \As updates \MM, only when both the end vertices belong to \MM. This leads to one of the vertices (say $v$) being removed from \MM, and the \cc of its neighbours being updated. Several of these neighbours can now have $count=0~$ and hence enter \MM. However, using Theorem~\ref{thm:simpleAlg} we know that the amortized time required by the update is only $O(deg(v))$, the degree of the vertex removed from \MM.

Our main idea is to modify \As to ensure that when an edge is inserted between two vertices in \MM, the end vertex with lower degree is removed from \MM. Note that in the original \As, the end vertex to be removed from MIS is chosen arbitrarily. We show that this key difference of choosing the lower degree (instead of arbitrary) end vertex to be removed from \MM, proves crucial in improving the amortized update time. We further prove our argument by showing worst case examples demonstrating the tightness of the upper bounds of the two algorithms (see Appendix~\ref{sec:tightInc}).

\subsubsection{Analysis}
We shall now analyze our incremental MIS algorithm. Consider the insertion of $i^{th}$ edge $(u_i,v_i)$, where without loss of generality $u_i$ is the lower degree vertex amongst $u_i$ and $v_i$. Using Theorem~\ref{thm:simpleAlg}, we know that the amortized update time of the algorithm is $O(deg(u_i))$ in case both $u_i,v_i\in$\MM, else $O(1)$. Hence, the total update time taken by the algorithm is $\sum_{i=1}^m O(deg(u_i))$.

We shall analyze the time taken by a vertex, during two {\em phases}, when it is {\em light} ($deg(u_i)\leq\sqrt{m}$), and when it becomes {\em heavy} ($deg(u_i)>\sqrt{m}$). If $u_i$ is light, then we  simply remove $u_i$ from \MM in $O(\sqrt m)$ amortized time (as $deg(u_i)\leq \sqrt m$).  If $u_i$ is heavy, then $v_i$ must also be heavy. However, there are only $O(\sqrt m)$ heavy vertices in the graph. Hence, a heavy vertex $u_i$ can be removed $O(\sqrt m)$ times from \MM\ due to its heavy neighbours. Further, the time taken for such a vertex is $O(deg_f(u_i))$, where $deg_f(u_i)(\geq deg(u_i))$ is the final degree of $u_i$ after the end of updates.     Thus, the total update time is calculated as follows.\\

\begin{tabular}{lllllll}
$\displaystyle\sum_{i=1}^m O(deg(u_i))$ & = & $\displaystyle\sum_{u_i~\text{is light}} O(deg(u_i))$ &+& $\displaystyle\sum_{u_i \text{is heavy}} O(deg(u_i))$\\

& $\le$ &$\displaystyle\sum_{i=1}^m O(\sqrt{m})$
& + & $\sum_{u \in V} |\{(u,v)\in E | v~ \text{is heavy}\}| \times O(deg(u))$\\

& $\le$ &$\displaystyle\sum_{i=1}^m O(\sqrt{m})$
& + & $\sum_{u \in V} \sqrt m \times O(deg_f(u))$ \\
& $\le$ &$O(m \sqrt m)$
& + & $O(m\sqrt m)$\\ \\
\end{tabular}

Hence, the total update time of our incremental algorithm is $O(m\sqrt{m})$. Also, since it is simply a special case of the simple MIS algorithm ~\cite{AssadiOSS18}, we also have adjustment complexity of amortized $O(1)$, and an upper bound of $O(m\Delta)$ total time, making the amortized update time $O(\min\{\Delta, \sqrt{m}\})$.


\newtheorem*{incMIS}{Theorem~\ref{thm:incMIS}}
\begin{incMIS}[Incremental MIS]
Given any graph $G=(V,E)$ having $n$ vertices and $m$ edges, MIS can be maintained in $O(\min\{\Delta,\sqrt{m}\})$ amortized time and $O(1)$ amortized adjustments per edge insertion, where $\Delta$ is the maximum degree of a vertex in $G$.
\end{incMIS}

\noindent
\textbf{Note: } This technique cannot be trivially extended to the fully dynamic setting. This is because here the crucial fact exploited is that the high degree vertices would be removed less number of times from MIS, which cannot be ensured in a fully dynamic environment. Also, as \As it can be trivially adapted to $\cal CONGEST$ model with amortized $O(1)$ rounds and adjustments per update. 

We have seen that the key difference of choosing the lower degree (instead of arbitrary) end vertex to be removed from \MM proves crucial in reducing the amortized update time. This argument can be proved by the following worst case examples demonstrating the tightness of the upper bounds of the two algorithms (see Appendix~\ref{sec:tightInc}). Note that this example for \As in incremental setting also shows tightness of the fully dynamic case, as the incremental setting is merely a special case of the fully dynamic setting.

\section{MIS with worst case guarantees}
\label{sec:wcMIS}
Assadi et al.~\cite{AssadiOSS18} demonstrated using a worst case example that the adjustment complexity (number of vertices which enter or leave \MM during an update) can be $\Omega(n)$ in the worst case. This justifies why the entire work on dynamic MIS is focused primarily on amortized bounds. However, in case we still want to achieve better worst case bounds, we are required to consider a {\em relaxed} model, where we settle at not maintaining the MIS {\em explicitly}. Rather we allow queries to answer whether a vertex is present in MIS after an update, such that the results of all the queries are consistent to some MIS of the graph.

\subsection*{Implicit Maintenance of MIS}

We now formally define this model for the dynamic maintenance of MIS. The model supports {\em updates} in the graph, such that the MIS is not {\em explicitly} maintained after each update. Additionally, the model allows queries of the form \IMv, which reports whether a vertex $v\in V$ is present in the MIS of the updated graph. Such a model essentially allows us to compute \MM partially, along with maintaining some  information which makes sure that the results of the queries are consistent with each other. Thus, this model allows us to achieve $o(n)$ time worst case bounds for both queries and updates, for the dynamic maintenance of MIS.

The underlying idea is as follows. Recall that in every update of \As exactly {\em one} vertex is removed from \MM and several vertices may be added to \MM (see Theorem~\ref{thm:simpleAlg}). Thus, it is easier to maintain an {\em independent set} rather than a {\em maximal independent set}, by always processing the removal of a vertex from \MM but not the insertion of vertices in \MM. Hence, we only make sure that after each update, no two vertices in \MM share an edge which leads to removing exactly one vertex from \MM. Now, whenever a vertex is queried we verify whether it is already in \MM or can be moved to \MM, and respond accordingly. In order to resolve these queries efficiently, we again maintain the {\em partial count} of neighbours of a vertex in \MM, which is described as follows.

We consider the vertices with high degree ($>\sqrt{m}$) as {\em heavy} and the rest as {\em light}, resulting in total $O(\sqrt{m})$ heavy vertices. Our algorithm maintains the {\em count} for only heavy vertices and not for light vertices. Thus, whenever a vertex enters or leaves \MM it only informs its heavy neighbours. We shall now describe the {\em update} and {\em query} algorithms.

The {\em update} algorithm merely updates the {\em count} of heavy end vertex (if any) in case of edge deletion, or edge insertion when both end vertices are not in \MM. For edge insertion having both end vertices in \MM, it removes one of the vertices from \MM and updates the {\em count} of its $O(\sqrt{m})$ heavy neighbours.
Note that the update algorithm does not ensure that \MM is an MIS, but necessarily ensures that \MM is an {\em independent set} since it always removes a vertex from \MM if any of its neighbours is in \MM.


A vertex $v$ may be added to \MM only if it is queried in \IMv as follows. If $v \in~$\MM, then we simply report it. Else, there are two cases depending on whether $v$ is light or heavy.  If $v$ is heavy and its {\em count} is zero, it implies that $v$ has no  neighbour in \MM. Hence, we simply add $v$ into \MM and update the {\em count} of its heavy neighbours.
However, if $v$ is light we do not have the {\em count} of its neighbours in \MM. Hence, we check if there is any neighbour of $v$ in \MM in $O(\sqrt m)$ time. If none of the neighbours of $v$ are in \MM, then we add $v$ to \MM and update the {\em count} of its heavy neighbours. This completes the query algorithm. It is easy to see that both {\em update} and {\em query} algorithms require $O(\sqrt{m})$ worst case time to add or remove at most one vertex from \MM and visit its $O(\sqrt{m})$ neighbours. Further, in case $\Delta< \sqrt{m}$, this update and query time reduces to $O(\Delta)$, as each vertex can have $O(\Delta)$ neighbours. 

Now, in the fully dynamic setting the value of $m$ may change significantly after sufficient updates. So we instead define the {\em heavy} status of a vertex using a constant $m_c$ which is initialized as $m_c=m$, and we ensure that $m_c/2<m<2m_c$. In case the value of $m$ increases beyond $2m_c$, we simply need to make some {\em heavy} vertices light by removing their corresponding value of {\em count}. This can be performed in a single step in $O(\sqrt{m})$ time as there are only $O(\sqrt{m})$ heavy vertices, and we update $m_c=2m_c$. However, in case the value of $m$ decreases below $m_c$, we need to compute the value of {\em count} for the {\em light} vertices in each update, which will become {\em heavy} if $m_c$ is reduced to half. Since total degree of all vertices is $O(m)$, there can be only $O(\sqrt{m})$ such vertices. Hence, in the next $O(\sqrt{m})$ updates we can compute the {\em count} for one such vertex in each update requiring $O(\sqrt{m})$ time. Since we have $O(m)$ updates before the value of $m=m_c/2$ and we update $m_c=m_c/2$, all the new {\em heavy} vertices will have computed its value of {\em count}. Also, if an edge update changes the {\em heavy} status of its end points, their corresponding {\em count} can be updated in $O(\sqrt{m})$ time.


In order to prove the correctness of our algorithm, we are required to prove that (1) \MM remains an independent set throughout the algorithm, and (2) All the neighbours of a vertex $v$ are verified before answering a query \IMv. The former clearly follows from the update algorithm, and to prove the latter we look at the heavy and light vertices separately. Since every vertex entering or leaving \MM necessarily informs its heavy neighbours, the second condition is clearly true for heavy vertices. For light vertices, the correctness of the second condition follows from the query algorithm. Finally, since at most one vertex (if any) leaves MIS in an update or enters MIS in a query, the adjustment complexity is $O(1)$. However, if we consider adjustment complexity considering updates only, it is still amortized $O(1)$ (similar to \As). Thus we have the following theorem. 

\begin{theorem}[Fully dynamic MIS (worst case)]
Given any graph $G=(V,E)$ having $n$ vertices and $m$ edges, MIS can be implicitly maintained under fully dynamic edge updates requiring $O(1)$ adjustments per update and query, allowing queries of the form \IMv, where both update and query requires worst case $O(\min\{\Delta,\sqrt{m}\})$ time.
\end{theorem}

\noindent
\textbf{Note:}
The above described algorithm can be trivially adapted to distributed $\cal CONGEST$ model requiring $O(\min\{\Delta,\sqrt{m}\})$ messages, and $O(1)$ rounds and adjustments per update or query in the worst case.\\

\noindent
\textbf{Remark:} The algorithm can also support vertex deletion similar to edge insertion in the same time. However, vertex insertion with an arbitrary number of incident edges would not be allowed as processing input itself may require $O(\Delta)$ time. Thus, allowing fully dynamic vertex updates requires $O(\Delta)$ worst case update time. However, in case the vertices are inserted without any incident edges the same complexity of $O(\min\{\Delta,\sqrt{m}\})$ is also applicable for fully dynamic vertex updates.

\section{Maximum Flow and Maximum Matching } 
A standard approach to both these problems uses the concept of {\em augmenting paths}, where finding an augmenting path allows us to increase the value of the solution by a single unit. Computing an augmenting path takes $O(m)$ time for both the problems and in the unweighted case the maximum value of the solution can be $O(n)$. This simply gives an $O(mn)$ time algorithm to compute the solution for both the problems in the static setting. 

However, maintaining them in the dynamic setting have not been explicitly examined. We report that both these bounds can be exactly matched by trivial extensions of the {\em augmenting path} algorithms in the incremental setting. Concretely, the incremental reachability algorithm~\cite{Italiano86} can be used to incrementally compute an augmenting path for maximum flow in total $O(m)$ time. Similarly, the blossoms algorithm~\cite{Gabow74} (which is inherently incremental) can be used to incrementally compute an augmenting path for maximum matching in $O(m)$ time. Once the augmenting path is found, the solution is updated and its value is increased by $1$ unit. Then we restart the computation of augmenting path in the updated graph. This process can be repeated $O(n)$ times, which is the maximum value of the solution. Hence, both the problems can be solved in total $O(mn)$ time, matching the lower bound by Dahlgaard~\cite{Dahlgaard16}. Furthermore, the fully dynamic algorithms for both these problems requiring $O(m)$ time per update are known as {\em folklore}, though not explicitly stated in the literature. Both these algorithms are also based on the {\em augmenting path} approach. We also state those algorithms for the sake of completeness. Refer to Appendix~\ref{sec:maxFlow} and Appendix~\ref{sec:mcMatch} for details.

\section{Conclusion}
We have presented several surprisingly simple algorithms for fundamental graph problems in the dynamic setting. These algorithms either improve the known upper bounds of the problem or match the known lower bounds. Additionally, we considered some relaxed settings under which such problems can be solved better. A common trait among all these algorithms is that they are extremely simple and use no complicated data structures, making it suitable for even classroom teaching of fundamental concepts as {\em amortization} and introducing {\em dynamic graph algorithms}. 

In the dynamic MIS problem, we also discuss the hardness of the problem in the dynamic setting. Most graph problems (as connectivity, reachability, maximum flow, maximum matching, MST, DFS, BFS etc.) are found to be harder to handle vertex updates instead of edge updates and handle deletions instead of insertions. This is surprisingly the opposite case with dynamic MIS, where except for edge insertions (which is the easiest update for most other problems), a trivial algorithm solves the problem optimally. Notably, this is also the case a few other fundamental problems as topological ordering, cycle detection, planarity testing, etc. We conjecture the reason for such a behaviour to be the following fundamental property: {\em If the solution of the problem is still valid (though sub-optimal) after an update, it shall be easier to handle the update}. This supports the behaviour of the problems mentioned above, both for which edge insertions are easiest, and for which they are hardest. 

Finally, in the light of the above discussion, we propose some future directions of research in these problems. It seems that the fully dynamic MIS under edge updates, shouldn't be much harder than the incremental setting. Hence, it would be interesting to see an algorithm to maintain fully dynamic MIS in $O(\min\{\Delta, \sqrt{m}\})$ amortized update time, which can preferably also be extended to the distributed $\cal CONGEST$ model with amortized $O(1)$ round and adjustment complexity. On the other hand, we believe decremental unit capacity maximum flow and maximum cardinality matching would be harder than the incremental setting. Hence, stronger lower bounds for these problems in the decremental setting would be interesting.


\bibliography{references}
\appendix
\section{Tightness of Incremental Algorithms}
\label{sec:tightInc}
We now present worst case examples demonstrating the tightness of the analysis of \As and our incremental MIS algorithm. These essentially highlight the difference between arbitrary removal and degree biased removal of an end vertex on insertion of an edge between two vertices of the MIS.

\subsection{Arbitrary removal}
\begin{figure}[ht]
\centering
\includegraphics[width=.3\linewidth]{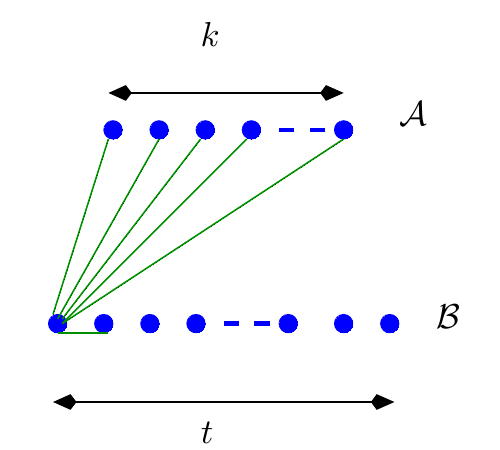}
\caption{Worst case example for arbitrary removal} 
\label{tightEg1}
 \end{figure}

We start with an empty graph where all the vertices are in MIS. Let the vertices be divided into two sets ${\cal A}=\{a_1,...,a_k\}$ and ${\cal B}=b_1,...,b_t$ (refer to Figure \ref{tightEg1}), where $\cal A$ has $k= m/\Delta$ vertices and $\cal B$ has the remaining $t=n-m/\Delta= O(n)$ vertices. 

We shall divide the insertion of edges into $t$ phases, where in the $j^{th}$ phase we add an edge from each vertex of $\cal A$ to $b_j$. And we always chose the vertex of $\cal A$ to be removed from MIS \MM. At the end of the phase, all vertices of $\cal A$ are out of \MM and are connected to $b_j\in$\MM. The phase ends with the addition of an edge between $b_j$ and $b_{j+1}$, which removes $b_j$ from \MM and hence all vertices of $\cal A$ are moved back to \MM. Since each vertex is allowed only $\Delta$ neighbours, and each phase adds a neighbour to each vertex in $\cal A$, we stop after $t^*=\Delta$ phases.

Hence, after $t^*$ phases we have added all the edges between ${\cal A}$ and the first $t^*$ vertices in ${\cal B}$, and among the first $t^*$ adjacent vertices of $\cal B$. Overall we add $O(|{\cal A}|\times t^*+t^*)$ edges, which equals to $O(kt^*)=O(\frac{m}{\Delta}*\Delta)=O(m)$ edges. 

Using Theorem~\ref{thm:simpleAlg}, the total edges processed during the $j^{th}$ phase, is the sum of the degrees of vertices that were removed from \MM, i.e., all the vertices in $\cal A$ and $b_j$. Now, in the $j^{th}$ phase the degree of each vertex in $\cal A$ is $j-1$, being connected to $b_1,..., b_{j-1}$ and the degree of $b_j$ is $k$. Hence, the total work in $j^{th}$ phase is $|{\cal A}|\times (j-1)+k = k\times j$. Thus, the total work done over all phases is
\begin{equation*}
	\sum_{j=1}^{t^*} \Omega(k\times j) = \Omega(k \times t^{*2}) = \Omega(\frac{m}{\Delta}\times \Delta^2) = \Omega(m\Delta)
\end{equation*}

Thus, we have the following bound for \As in incremental (and hence fully dynamic) setting.

\begin{theorem}
For each value of $n\leq m\leq {n\choose 2}$ and $1\leq \Delta \leq n$, there exists a sequence of $m$ edge insertions where the degree of each vertex is bounded by $\Delta$ for which \As requires total $\Theta(m\Delta)$ time to maintain the MIS.
\label{thm:wcS}
\end{theorem}

\subsection{Degree biased removal}
In this example, we consider our incremental MIS algorithm, which chooses the end vertex with the lower degree to be removed from \MM when an edge is inserted between two vertices in \MM. We essentially modify the previous example to make sure the vertices of $\cal A$ necessarily fall when connected to the vertices in $\cal B$. 

\begin{figure}[ht]
\centering
\includegraphics[width=.45\linewidth]{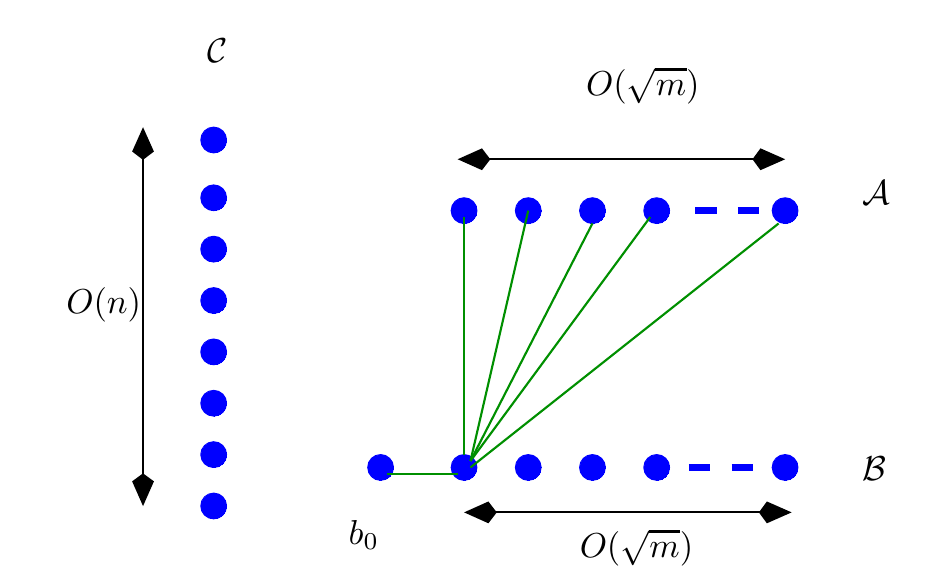}
\caption{Tightness Example for degree biased removal} 
\label{tightEg2}
 \end{figure}

Let the vertices to be divided into two sets ${\cal A}=\{a_1,...,a_k\}$ and ${\cal B}=\{b_1,...,b_t\}$ as before, and an additional set $\cal C$ of residual vertices to ensure that degrees of vertices in $\cal B$ are sufficiently high (refer to Figure \ref{tightEg2}), where $k=t=\sqrt{m}/4$. We connect each vertex $b_i$ with some $\sqrt{m}+1$ vertices in $\cal C$. Additionally, we have a vertex $b_0$, connected to all the $O(n)$ vertices in $\cal C$. We initialize the MIS with all vertices of $\cal A$,  $\cal B$ and $b_0$ in \MM.

Again, we divide the insertion of edges into $t$ phases, where in the $j^{th}$ phase we add an edge from each vertex of $\cal A$ to $b_j$. Since the maximum degree of a vertex in ${\cal A}$ is $|\cal B|$, and degree of each $b_j$ is $\sqrt{m}+1$, we always have the vertex of $\cal A$ to be removed from MIS \MM. At the end of the phase, all vertices of $\cal A$ are out of \MM and are connected to $b_j\in$\MM. The phase ends with the addition of $(b_j,b_{0})$, which removes $b_j$ from \MM and hence all vertices of $\cal A$ are moved back to \MM. 

Hence, after $t$ phases we have added all the edges between ${\cal A}$ and ${\cal B}$, between each vertex of $\cal B$ and $b_0$. The initial graph already had each vertex of $\cal B$ connected to some $\sqrt{m}+1$ neighbours in $\cal C$ and $b_0$ connected to all neighbours of $\cal C$. Overall we add $O(|{\cal A}|\times |{\cal B}|+ |{\cal B}|\times (\sqrt{m}+1)+{|\cal B|}+n)$ edges, which equals to $O(kt+t\sqrt{m}+n)=O(m+n)$ edges.

Again, using Theorem~\ref{thm:simpleAlg}, the total edges processed during the $j^{th}$ phase, is the sum of the degrees of vertices that were removed from \MM, i.e., all the vertices in $\cal A$ and $b_j$. Now, in the $j^{th}$ phase the degree of each vertex in $\cal A$ is $j-1$, being connected to $b_1,..., b_{j-1}$ and the degree of $b_j$ is $\Omega(\sqrt{m})$. Hence, the total work done in $j^{th}$ phase is $\Omega(|{\cal A}|\times j+\sqrt{m}) = \Omega(k\times j+\sqrt{m})$. Thus, the total work done over all phases is
\begin{equation*}
	\sum_{j=1}^{t} \Omega(k\times (j-1) + \sqrt{m}) = \Omega(k \times t^2+k\sqrt{m}) = \Omega(\sqrt{m}\times \sqrt{m}^2 + m) = \Omega(m\sqrt{m})
\end{equation*}

Thus, we have the following bound for our incremental MIS algorithm.

\begin{theorem}
For each value of $n\leq m\leq {n\choose 2}$, there exists a sequence of $m$ edge insertions for which our incremental algorithm requires total $\Theta(m\sqrt{m})$ time to maintain the MIS.
\label{thm:wcI}
\end{theorem}

\noindent
\textbf{Remark: } This example is similar to the previous example with an exception that in this case, the degree of vertices in $\cal B$ should remain higher than the degree of a vertex in $\cal A$, i.e., $|\cal B|$ at the end of all the phases. Hence, $|{\cal B}|\leq \sqrt{m}$ else the total edges in $G$ would be $\Omega(m)$. Thus, the number of neighbours of  $\cal A$ in $\cal B$ cannot be increased despite choosing any large value of $\Delta$. As evident from the previous example in absence of such a restriction, the amortized time can be raised to $\Delta$ implying the significance of the degree biased removal in the incremental algorithm.

\section{Unit capacity Maximum Flow}
\label{sec:maxFlow}
The maximum flow problem is one of the most studied combinatorial optimization problem having a lot of practical applications (see \cite{GoldbergT14} for a survey). Given a graph $G=(V,E)$ having $n$ vertices and $m$ edges where each edge has a capacity (positive real weight) $c:E\rightarrow \mathbb{R^+}$ associated to it. A flow from a {\em source} $s$ to a {\em sink} $t$ is an assignment of flow $f:E\rightarrow \mathbb{R^+}$ to each edge such that it satisfies the following two constraints. Firstly, the {\em capacity} constraints imply that the flow on each edge is limited by its capacity, i.e. $\forall e\in E, f(e)\leq c(e)$. Secondly, the {\em conservation} constraints imply that for each non-terminal vertex ($v\in V\setminus\{s,t\}$), the flow passing through $v$ is conserved, i.e. $\sum_{(u,v)\in E} f(u,v)=\sum_{(v,w)\in E} f(v,w)$. The amount of flow leaving $s$ or entering $t$ is referred as the flow $F$ of the network, i.e., $F=\sum_{(s,u)\in E} f(s,u)=\sum_{(v,t)\in E} f(v,t)$. Clearly, in a unit capacity simple graph $F=O(n)$, corresponding to each edge leaving $s$ (or entering $t$). The maximum flow problem evaluates the flow $f$ from $s$ to $t$ which maximizes the value of $F$. 

Orlin \cite{Orlin13} presented an algorithm to find the maximum flow in $O(mn)$ time. For integral arc capacities (bounded by $U$), it can be evaluated $O(m\cdot \min(n^{2/3}, m^{1/2}) \log(n^2/m) \log U))$ time~\cite{GoldbergR98}, which further reduces to $O(m\cdot \min(n^{2/3}, m^{1/2}))$ time for unweighted (unit capacity) graphs~\cite{EvenT75}. In the dynamic setting, a few algorithms are known for maintaining maximum flow. Kumar and Gupta~\cite{KumarG03} showed that partially dynamic maximum flow can be maintained in $O(\Delta n^2 m)$ update time, where $\Delta n$ is the number of vertices whose flow is affected. Other algorithms are by Goldberg et al.~\cite{GoldbergHKKTW15} and Kohli and Torr~\cite{KohliT10}, which work very well in practice on computer vision problems but do not have strong asymptotic guarantees.

Recently, Dahlgaard \cite{Dahlgaard16} proved conditional lower bounds for partially dynamic problems including maximum flow  under OMv conjecture. For directed and weighted sparse graphs, no algorithm can maintain partially dynamic max flow in $O(m^{1-\epsilon})$ amortized update time. For directed unweighted (unit capacity) graphs and undirected weighted graphs, no algorithm can maintain partially dynamic maximum flow in $O(n^{1-\epsilon})$ amortized update time. We report a trivial extension of the incremental reachability alogrithm~\cite{Italiano86} to solve the incremental unit capacity (unweighted) maximum flow problem matching its corresponding lower bound.

We shall now describe some simple dynamic algorithms for maintaining maximum flow in unit capacity graphs. The main idea behind such algorithms is based on the classical {\em augmenting path} approach of the standard Ford Fulkerson~\cite{FordF62} algorithm. In the interest of completeness, before describing our incremental algorithm for maintaining maximum flow in $O(F)$ (where $F=O(n)$) amortized time per update, we present the {\em folklore} algorithm supporting fully dynamic edges updates in $O(m)$ time. To describe these algorithms succinctly, we briefly describe the {\em augmenting path} approach and the concept of {\em residual graphs}.

\subsubsection*{Residual Graphs and Augmenting paths}
Given an unweighted (unit capacity) directed graph $G=(V,E)$, having flow function defined on each edge $f:E\rightarrow \{0,1\}$. The {\em residual graph} is computed by changing the direction of every edge $e$, if $f(e)=1$. Note that this may create two edges in a graph between the same endpoints. This residual graph allows a simple characterization of whether the current flow $F$ is indeed a maximum flow because of the following property: {\em If there exists an $s-t$ path in the residual graph, referred as {\em augmenting path}, the value of the flow can be increased by pushing flow along the {\em augmenting} path in the residual graph}~\cite{FordF62}. This results in reversing the direction of each edge on this path in the updated residual path. The flow reaches the maximum value when there does not exist any augmenting path, i.e. $s-t$ path, in the residual graph.

\subsection{Fully Dynamic Maximum Flow}
We now describe the {\em folklore} algorithm for maintaining Maximum Flow of a unit capacity graph under fully dynamic edge updates. The flow is updated by the computation of a single augmenting path in the residual graph after the corresponding edge update as follows.

\begin{itemize}
\item \textbf{Insertion of an edge:}  An edge insertion can either increase the maximum flow by one unit or leave it unchanged. Recall that when $f$ is a maximum flow, the sink $t$ is not reachable from the source $s$ in the residual graph. Hence if the inserted edge creates an $s-t$ path in the residual graph, the maximum flow increases by exactly one unit. Thus, the algorithm finds an $s-t$ path in the residual graph and pushes a flow of one unit along the path. If no such path is found, the maximum flow has not increased and the solution remains unchanged.


\item \textbf{Deletion of an edge:} If the deleted edge doesn't carry any flow, the flow remains unchanged. Otherwise, the edge deletion can either decrease the maximum flow or simply make the current flow invalid requiring us to reroute the flow without changing its value. Hence, on deletion of an edge $(x,y)$ the algorithm first attempts to restore the flow by finding an alternate path from $x$ to $y$ in the residual graph. If such a path is found the algorithm pushes a flow of unit capacity along the path, restoring the maximum flow in the graph. Otherwise, we have to send back one unit of flow each, from $x$ to $s$ and from $t$ to $y$, to reduce the maximum flow by one unit. For this an edge $(s,t)$ is added and a path from $x$ to $y$ is found,  which necessarily exists containing the edge $(s,t)$. Again, the algorithm pushes a flow of unit capacity along the path, restoring the maximum flow in the graph. After updating the residual graph, we remove the extra added edge $(s,t)$ from the graph.
\end{itemize}

Thus, each update can be performed in $O(m)$ time, as it performs $O(1)$ reachability queries (using BFS/DFS traversals) and updates the residual graph.
Hence, we have the following theorem.

\begin{theorem}[Fully dynamic Unit Capacity Maximum Flow (folklore)]
Given an unweighted (unit-capacity) graph $G=(V,E)$ having $n$ vertices and $m$ edges, fully dynamic maximum flow under edge updates can be maintained in $O(m)$ worst case time per update.
\end{theorem}

\subsection{Incremental Maximum Flow}
In the incremental setting, the unit-capacity maximum flow can be maintained using amortized $O(F)$ update time. Recall that in the fully dynamic algorithm, on insertion of an edge the flow is increased only if the $t$ becomes reachable from $s$ in the residual graph as a result of the update. Trivially, verifying whether $t$ is reachable after each update requires $O(m)$ time per update, even when the flow is not increased. However, this can be computed more efficiently using the single source incremental reachability algorithm~\cite{Italiano86} requiring total $O(m)$ time for every increase in the value of flow. In the interest of completeness, let us briefly describe the incremental reachability algorithm~\cite{Italiano86} as follows.

\subsubsection*{Single source incremental reachability~\cite{Italiano86}}
This algorithm essentially maintains a reachability tree $T$ from the source vertex $s$. This tree $T$ is initialized with a single node $s$ and grown to include all the vertices reachable from $s$. On insertion of an edge $(u,v)$, an update is required only when $u\in T$ and $v\notin T$. In such a case, the edge $(u,v)$ is added making $v$ a child of $u$. Further, the update algorithm processes every outgoing edge $(v,w)$ of $v$, i.e., to find vertices $w\notin T$, which are added to $T$ recursively using the same procedure. Thus, this process continues until all the vertices reachable from $s$ are added to $T$. Clearly, the process takes total $O(m)$ time as each edge $(u,v)$ is processed $O(1)$ times, when it is inserted in the graph and when $u$ is added to $T$.

\subsubsection*{Algorithm}
The algorithm prominently uses the fact that the value of maximum flow $F$ in a unit capacity simple graph is $O(n)$ since the number of outgoing edges from $s$ (or the incoming edges to $t$) is $O(n)$. Our algorithm is divided into $F$ stages where at the end of each stage the value of maximum flow increases by a single unit. Each stage starts by building an incremental single source reachability structure~\cite{Italiano86}, i.e. the reachability tree $T$, from $s$ on the residual graph. On insertion of an edge, the reachability tree $T$ is updated using the incremental reachability algorithm. The stage continues until $t$ becomes reachable from $s$ and added to $T$. This gives the $s-t$ augmenting path and we push a unit of flow along the path and update the residual graph. Thus, each stage requires total $O(m)$ time for maintaining incremental reachability structure~\cite{Italiano86}, taking overall $O(mF)$ time (where $F=O(n)$), giving us the following result.

\begin{theorem}[Incremental Unit Capacity Maximum Flow]
Given an unweighted (unit-capacity) graph $G=(V,E)$ having $n$ vertices and $m$ edges, incremental maximum flow can be maintained in amortized $O(F)$ update time, where $F=O(n)$ is the value of the maximum flow of the final graph.
\label{thm:incMF}
\end{theorem}


\noindent
\textbf{Remark: } At the end of a stage, it is necessary to rebuild the incremental reachability structure~\cite{Italiano86} from scratch as it does not support edge deletions or reversals. This is required because when the flow is pushed along the $s-t$ path the residual graph is updated by reversing the direction of each edge on the $s-t$ path.

\section{Maximum Cardinality Matching}
\label{sec:mcMatch}
Maximum Matching is one of the most prominently studied combinatorial graph problems having a lot of practical applications. For a given graph $G=(V,E)$ with $n$ vertices and $m$ edges, a set of edges $\Em\subseteq E$ is called a {\em matching}, if no two edges in \EM share an end vertex in $V$. In the Maximum Matching problem, the aim is to compute the matching \EM of the maximum weight. In case the graph is unweighted, the problem is called Maximum Cardinality Matching. Micali and Vazirani presented an algorithm to compute maximum cardinality matching in $O(m\sqrt{n})$ time. For the weighted case, the fastest algorithm is by Galil et. al~\cite{GalilMG86} requiring $O(mn\log n)$ time which improves the $O(n^3)$ time algorithm~\cite{Gabow74} for sparse graphs. In the dynamic setting, only the problem of computing $\alpha$-approximate matching (having cardinality $\geq\Em/\alpha$) has been extensively studied~\cite{Solomon16,Bhattacharya2016,Bernstein2016a}.

%

Recently, Dahlgaard \cite{Dahlgaard16} proved conditional lower bounds for partially dynamic problems including maximum matching under OMv conjecture. For bipartite graphs (and hence general graphs), no algorithm can maintain partially dynamic maximum cardinality matching in $O(n^{1-\epsilon})$ amortized update time. We report a trivial extension of the classical blossom's algorithm~\cite{Edmonds87,Gabow74}, to solve this problem for general unweighted graphs matching its corresponding lower bound.

We first like to point out that maximum bipartite matching can be easily solved by a maximum flow algorithm using the standard reduction~\cite{KleinbergT05}. Hence an incremental algorithm for maintaining unit capacity maximum flow, also solves the problem of maintaining incremental maximum matching in bipartite graphs in the same bounds, i.e., amortized $O(n)$ update time. This matches the lower bound of $\Omega(n)$ given by Dahlgard~\cite{Dahlgaard16} for bipartite matching. In this section, we show that the same upper bound can also be maintained for general graphs by using a {\em trivial} extension of the Blossoms algorithm~\cite{Edmonds87,Gabow74} for finding an augmenting path, which is defined differently for maximum matching as follows.

\subsubsection*{Augmenting paths}
Given any graph $G=(V,E)$, a matching $\Em\subseteq E$ is a set of edges such that no two edges in \EM share an endpoint. The vertices on which some edge from \EM (also called {\em matched} edge) is incident is called a {\em matched} vertex. The remaining unmatched vertices are called {\em free} vertices. A given matching \EM is called the maximum matching if there does not exist an {\em augmenting path}, which is defined as follows: A {\em simple} path which starts and ends at a free vertex such that every odd edge on the path is a {\em matched} edge, and hence all the intermediate vertices are {\em matched} vertices. In case such a path $p$ exists, it can be used to increase the cardinality of the matching \EM, by removing all the {\em matched} edges on $p$ from \EM and adding all the {\em unmatched} edges on $p$ to \EM.

\subsection{Fully Dynamic Maximum Cardinality Matching}
\label{sec:fdMCM}
Computing the augmenting path starting from a free vertex $v$ requires a single BFS traversal from $v$ requiring $O(m)$ time, where on even layers the algorithm only explores {\em matched} edges. This results in trivial {\em folklore} fully dynamic maximum matching algorithm requiring $O(m)$ worst case update time as follows.

The algorithm essentially selects the end vertex $v$ based on the update. In case of vertex insertion, the inserted vertex is chosen as $v$. In case of deletion of a vertex $x$, \EM is not updated in case $x$ was a {\em free} vertex. Else if $x$ was matched to $y$, we start the computation of augmenting path from $y$ (as $v$). In case of insertion of an edge $(x,y)$, \EM is not updated if both $x$ and $y$ are {\em matched}. Else, if both are {\em free} we simply add $(x,y)$ to \EM. Else the augmenting path is computed starting from the {\em free} vertex amongst $x$ and $y$ (as $v$). Finally, in case of deletion of an edge $(x,y)$, \EM is not updated if $(x,y)$ was not a matched edge. Else, the augmenting path is computed twice starting from $x$ and $y$ (as $v$) in the two computations. Thus, each update can be performed in $O(m)$ time resulting in the following theorem.

\begin{theorem}[Fully dynamic Maximum Cardinality Matching (folklore)]
Given a graph $G=(V,E)$ having $n$ vertices and $m$ edges, fully dynamic maximum cardinality matching under edge or vertex updates can be maintained in $O(m)$ worst case time per update.
\end{theorem}

\subsection{Incremental Maximum Cardinality Matching}
In the incremental setting Maximum Cardinality Matching can be maintained in $O(|\Em|)$ amortized time per update, where $|\Em|=O(n)$. Note that it does not directly follow from the fully dynamic case, as we don't know the corresponding starting vertex of the augmenting path. However, for computing {\em augmenting paths} from any starting vertex in $O(m)$ time, we can use the standard Blossom's algorithm~\cite{Edmonds87,Gabow74}. It turns out that the Blossom's algorithm trivially extends to the incremental setting, such that it requires $O(m)$ time per increase in the cardinality of maximum matching. In the interest of completeness, we briefly describe the Blossom's algorithm as follows.

\subsubsection*{Blossom's Augmenting Path algorithm~\cite{Edmonds87,Gabow74}}
The algorithm essentially maintains a tree from each free vertex, where each node is either a single vertex or a set of vertices which occur in form of a {\em blossom}. Further, it ensures that each vertex of the graph occurs in only one such tree. The vertices are called as {\em odd} or {\em even} if they occur respectively on the odd level  or the even level (including free vertices in level {\em zero}) of any tree, and {\em unvisited} otherwise. The children of {\em even} vertices are connected to them through {\em unmatched} edges, whereas those of {\em odd} vertices are connected through {\em matched} edges. Thus, all the paths from roots to leaves in such trees have alternating {\em unmatched} and {\em matched} edges, as required by an augmenting path.

The algorithm starts with all free vertices as {\em even} vertices on singleton trees, and the remaining vertices as {\em unvisited}. Then each edge $(u,v)$ is considered one by one (and hence naturally extends to incremental setting) updating the trees accordingly as follows: If no end point is {\em even}, ignore the edge. Else, without loss of generality let $u$ be an {\em even} vertex. If $v$ is an {\em unvisited} vertex, simply add $v$ as a child of $u$. Else, if $v$ is an {\em odd} vertex, simply ignore it. However, if $v$ is an {\em even} vertex, the action depends on the trees to which $u$ and $v$ belong. In case they belong to the same tree we get a {\em blossom}, which is processed as follows. Find the lowest common ancestor $w$ of $u$ and $v$, and shrink all the vertices on tree paths connecting $w$ with $u$ and $v$ to a single {\em blossom} vertex $w'$, to be placed in place of $w$. The children of the {\em even} vertices in this blossom, which are not a part of this blossom, become the children of $w'$. Also, now each {\em odd} vertex in this blossom, have become {\em even} so all its edges that were previously ignored are explored recursively using the same procedure. Note that the compressed blossoms need to be maintained using modified Disjoint Set Union algorithm~\cite{GabowT83} (taking overall $O(m)$ time), allowing a vertex $v$ to quickly identify the blossom (and hence node $w'$ in the tree) they belong to. Thus, an insertion of $(u,v)$ is first evaluated to find nodes (vertices or blossoms) of the final tree $u'$ and $v'$ representing $u$ and $v$, and the edge $(u',v')$ is processed accordingly.

Finally, in case both $u$ and $v$ are {\em even} and they belong to different trees, and we have found an augmenting path from the root of the first tree to $u$, through $(u,v)$, followed by the tree path from $v$ to its root. However, some vertices on this path may be the {\em shrinked blossoms} and hence to report the augmenting path they are needed to be {\em un-shrinked}. An alternate simpler way is to simply start the augmenting path computation using the simple algorithm from the root of the tree containing $v$, as it is necessarily one end vertex of the augmenting path. Thus, both the {\em blossom's} algorithm and augmenting path reporting can be performed in total $O(m)$ time.

\subsubsection*{Algorithm}
The algorithm prominently uses the fact that the maximum cardinality of a matching $|\Em|$ is $O(n)$, since the each vertex can have at most one {\em matched} edge incident to it. The algorithm is again divided into $|\Em|$ stages, where at the end of each stage the Blossom's algorithm detects an {\em augmenting path} and hence increases the cardinality of \EM by a single unit. Each stage starts by initiating the Blossom's algorithm~\cite{Edmonds87,Gabow74} from scratch in the updated matching, and continues inserting edges until the algorithm detects the augmenting path. Thus, each stage requires total $O(m)$ time for computing the augmenting path and updating \EM, taking overall $O(m|\Em|)$ time where $|\Em|=O(n)$ is the value of the maximum cardinality matching of the graph, giving us the following result.

\begin{theorem}[Incremental Maximum Cardinality Matching]
Given a graph $G=(V,E)$ having $n$ vertices and $m$ edges, incremental maximum cardinality matching can be maintained in amortized $O(|\Em|)$ update time, where $|\Em|=O(n)$ is the size of the maximum matching of the final graph.
\label{thm:incMCM}
\end{theorem}


\noindent
\textbf{Remark: } The algorithm also works for the case of incremental vertex updates. Also, Blossom's algorithm can also be used for fully dynamic updates having $O(m)$ update time. However, the algorithm described in Section~\ref{sec:fdMCM} is simpler and more intuitive.

\end{document}